\documentclass[a4paper]{article}

\usepackage{icrc2013}

\usepackage{graphicx}
\usepackage[english]{babel}
\usepackage{hyperref}

\newcommand{\fermi}{\textit{Fermi}}

\newcommand{\veritas}{VERITAS}

\newcommand{\aap}{A \& A}

\newcommand{\araa}{AR A \& A}
\newcommand{\apj}{ApJ }
\newcommand{\apjl}{ApJL}
\newcommand{\apjs}{ApJS}

\newcommand{\pasp}{PASP}

\title{Upper Limits From Five Years of VERITAS Blazar Observations}

\shorttitle{Upper limits from five years of VERITAS blazar observations}

\authors{
Matteo Cerruti $^{1}$,
for the VERITAS Collaboration.
}

\afiliations{
$^1$ Harvard-Smithsonian Center for Astrophysics; 60 Garden Street, Cambridge, MA 02138, USA \\
}

\email{matteo.cerruti@cfa.harvard.edu}

\abstract{The VERITAS array of Cherenkov telescopes was used to observe $\approx 130$ blazars from 2007 to 2012. Of these, 25 were detected as very-high-energy (VHE; E $>$ 100 GeV) sources. We present here the results of the analysis of 65 VERITAS non-detected blazars, including upper limits on their VHE flux. Results from a stacked analysis of the entire data set and of smaller sub-sets (defined as a function of the redshift and the blazar class) are presented and discussed.}


\begin{document}
\maketitle


\section{Introduction}
The extra-galactic sky observed at very-high-energies (VHE; E $>$ 100 GeV) is dominated by blazars, a class of active galactic nuclei (AGN). In the framework of the AGN unified model \cite{Urry}, blazars are considered as radio-loud AGN whose relativistic jet is aligned to the observer's line of sight. The observational properties of blazars are an optical/UV spectrum dominated by a non-thermal continuum, a high degree of polarization, extreme temporal variability and a spectral energy distribution (SED) composed of two distinct bumps, peaking in millimiter-to-X-rays and $\gamma$-rays, respectively. The blazar class is divided into the two sub-classes of FSRQs (flat-spectrum radio quasar) and BL Lac objects, depending on whether emission lines are observed (in FSRQs) or not (in BL Lacs) in the optical spectrum \cite{Angel}.\\

 The origin of the lower-frequency SED component is ascribed to synchrotron emission by a population of non-thermal electrons in the jet. The high-energy component is generally associated, in leptonic models, with inverse-Compton scattering of low energy photons (the synchrotron photons themselves, or an external photon field, see \cite{Konigl,Dermer}) off the electrons in the emitting region. The frequency of the synchrotron peak is used to further classify blazars (see e.g. \cite{Fermiagn}): if $\nu_{peak} < 10^{14}$ Hz, the object is classified as low-synchrotron-peaked blazar (LSP), while if $\nu_{peak} > 10^{15}$ Hz, it is classified as high-synchrotron peaked blazar (HSP). Sources with $10^{14} \leq \nu_{peak} \leq 10^{15}$ Hz are classified as intermediate-synchrotron-peaked blazars (ISP). While FSRQs are essentially all LSPs, BL Lac objects show a variety of synchrotron peak frequencies.\\
 Interestingly, the population of known VHE blazars is not homogeneous, being dominated by HSP blazars\footnote{see http://tevcat.uchicago.edu}. Another characteristic of VHE blazars is related to absorption on the extra-galactic background light (EBL, see \cite{Salamon}). VHE photons interact with EBL photons to pair-produce electron-positron pairs. This absorption effect defines a horizon, above which the VHE flux from the source is significantly reduced. Up to now, the most distant VHE blazars detected have a redshift of $z \approx$0.6 (see \cite{Amy, Kuv}).\\  

The \veritas\ array has observed more than one hundred blazars since its beginning of operation in 2007. Among them, only 25 have been detected as VHE emitters (using the standard significance threshold of $\sigma > 5$). We present here the preliminary results of the analysis of 65 non-detected blazars. The sample includes different blazar sub-classes (HSP vs ISP/LSP blazars) in different redshifts, several unidentified \fermi\ sources as well as two Seyfert 1 galaxies (type-1 radio-quiet AGN).  \\

\begin{figure*}[hbtp!]
\includegraphics[width=220pt]{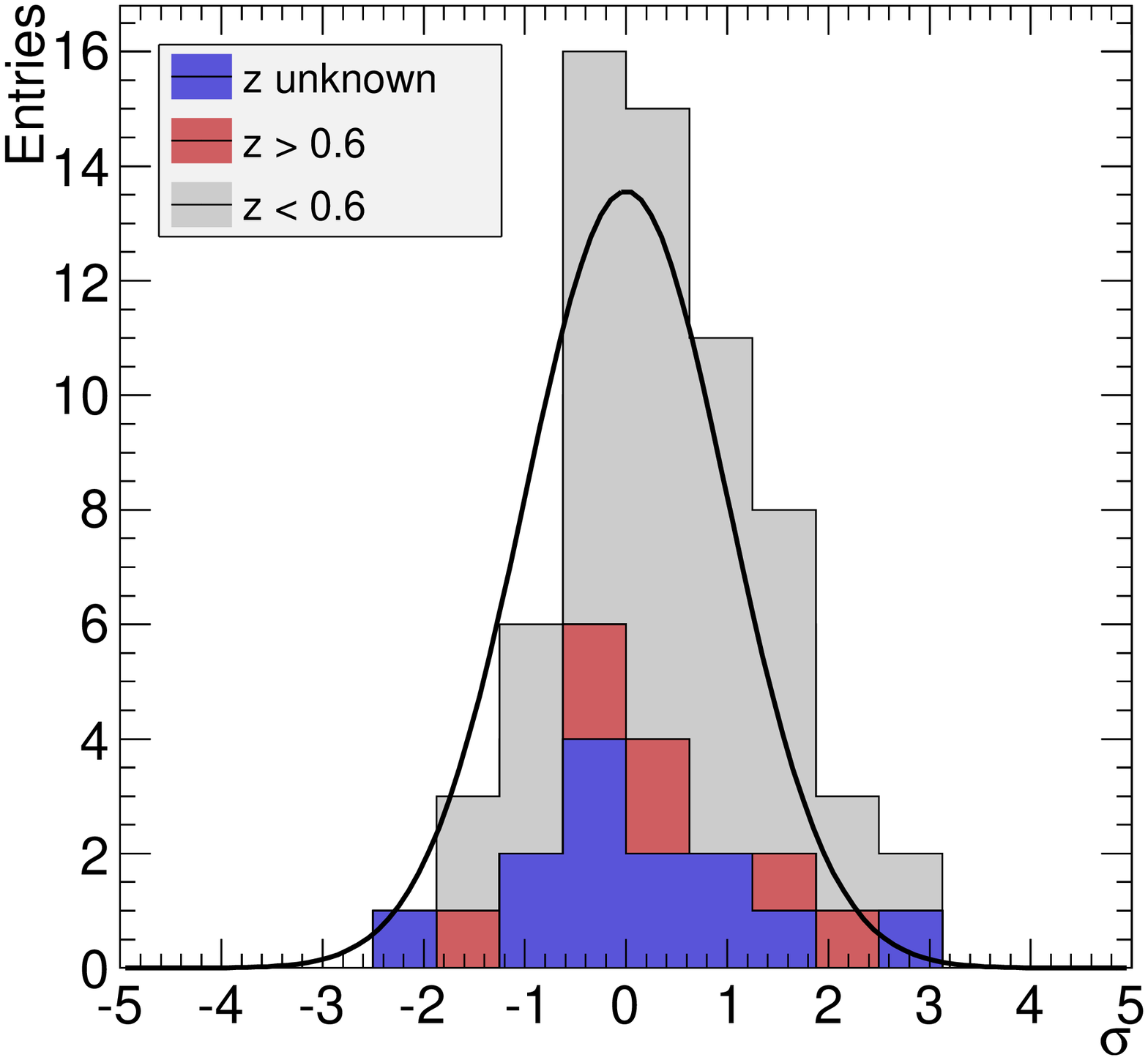}
\includegraphics[width=220pt]{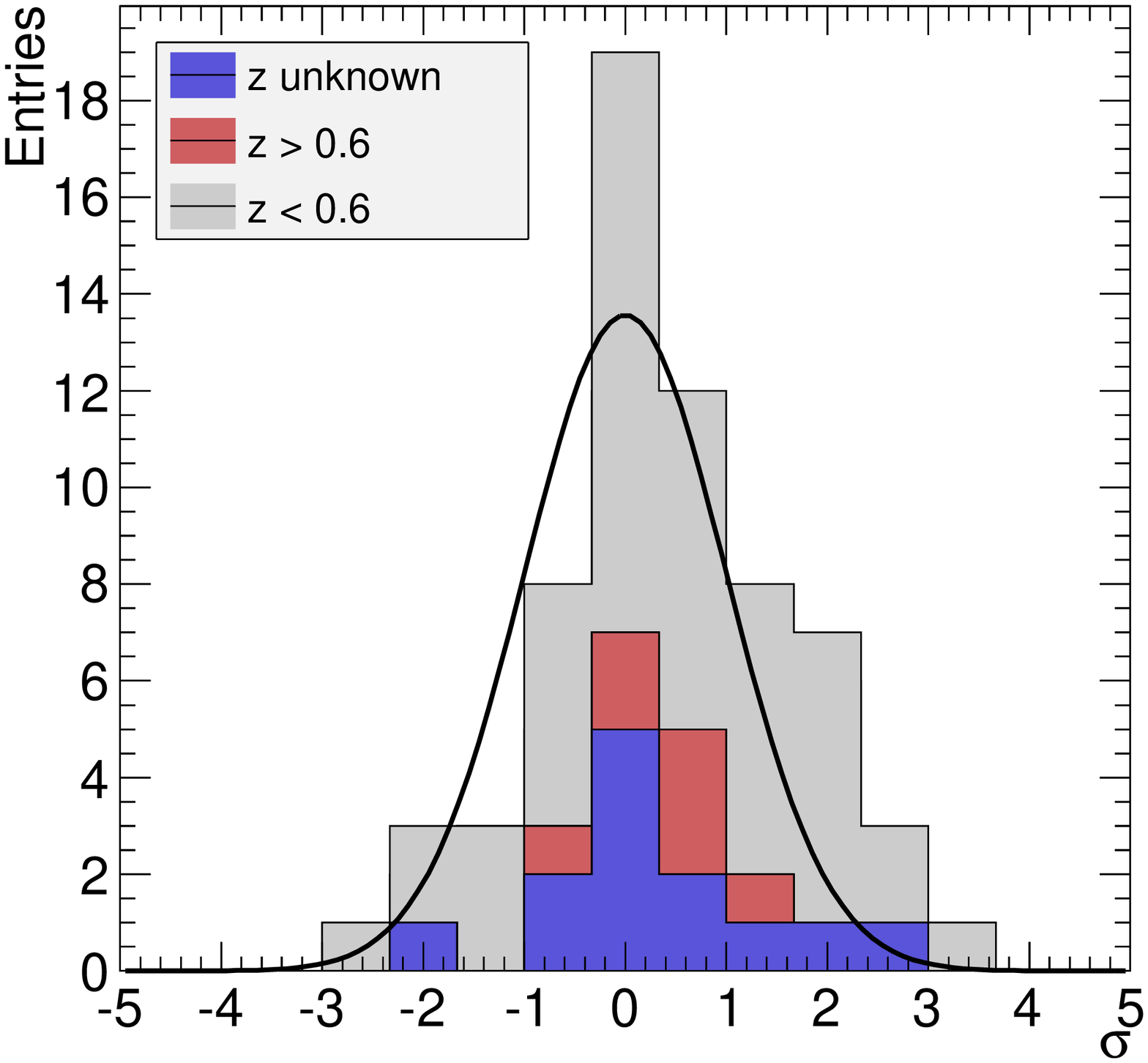}

\includegraphics[width=220pt]{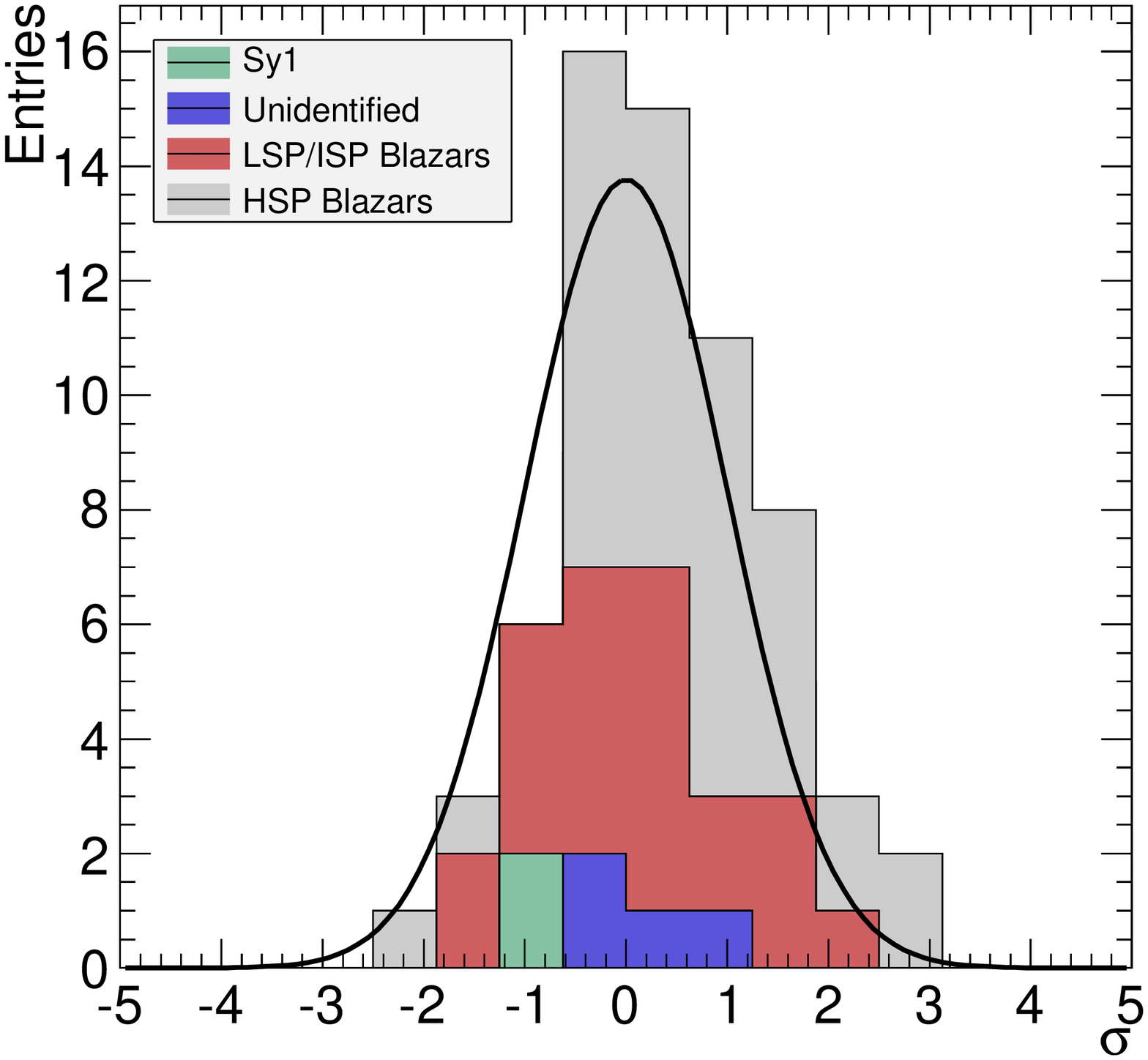}
\includegraphics[width=220pt]{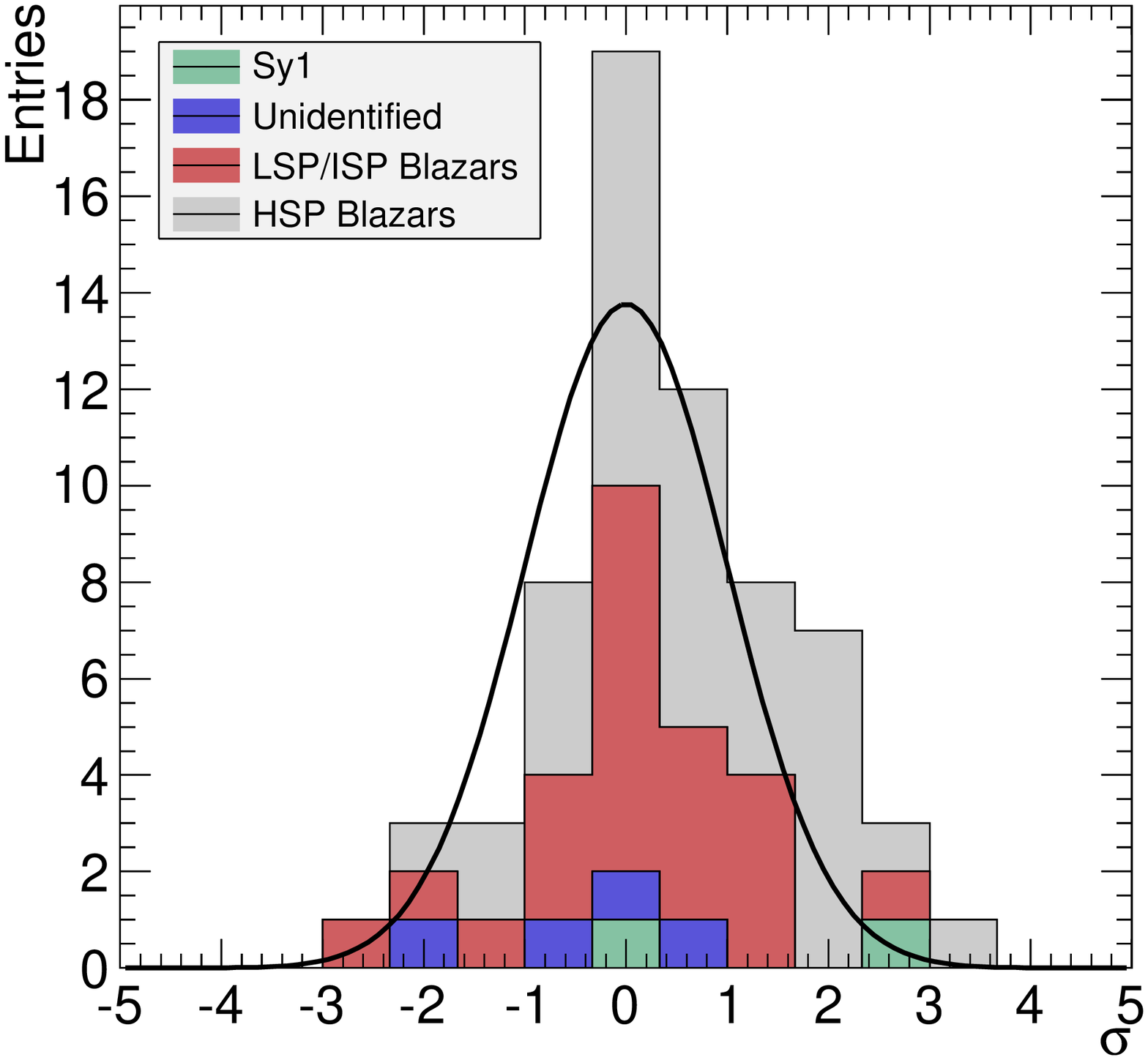}

\includegraphics[width=220pt]{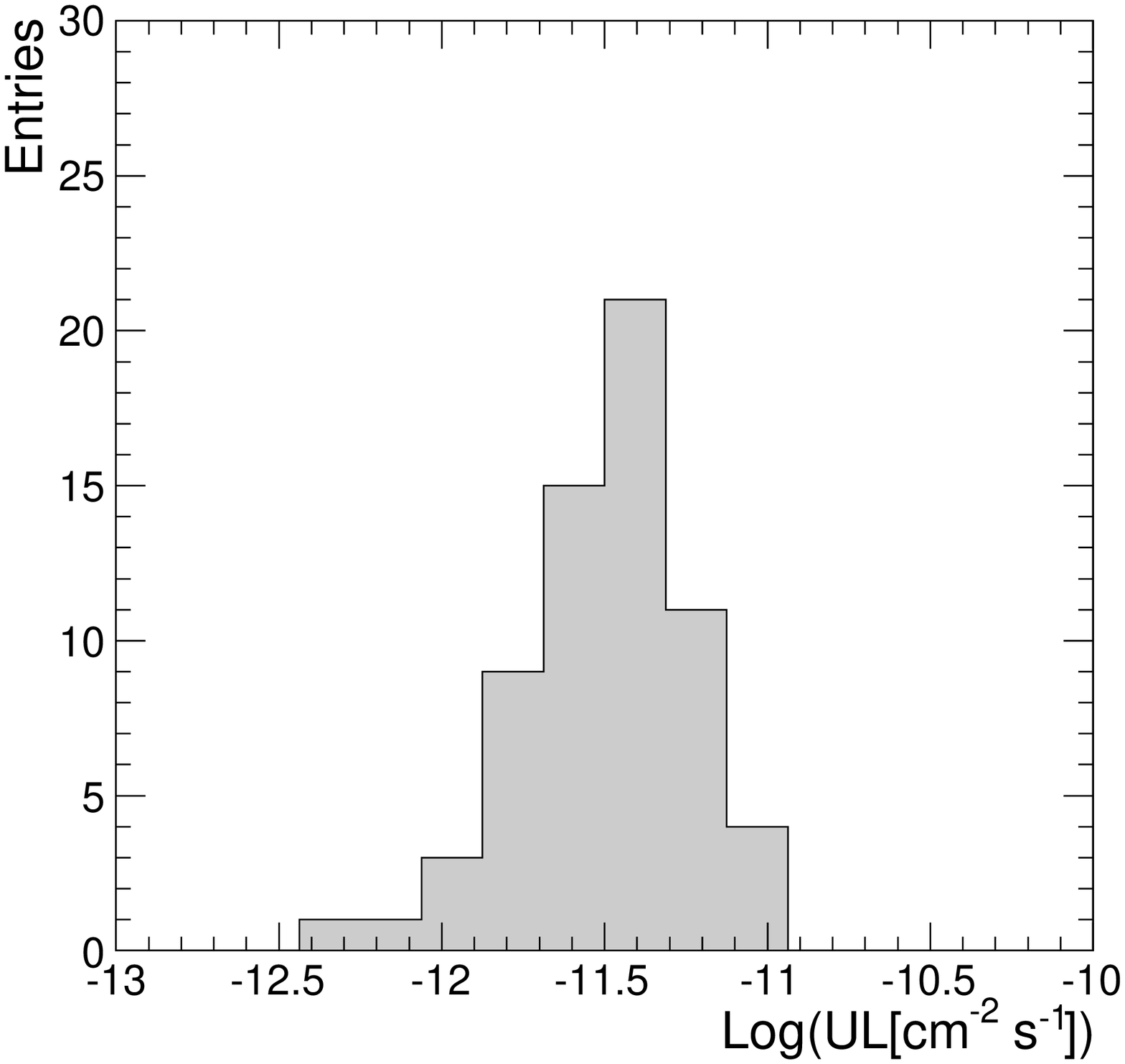}
\includegraphics[width=220pt]{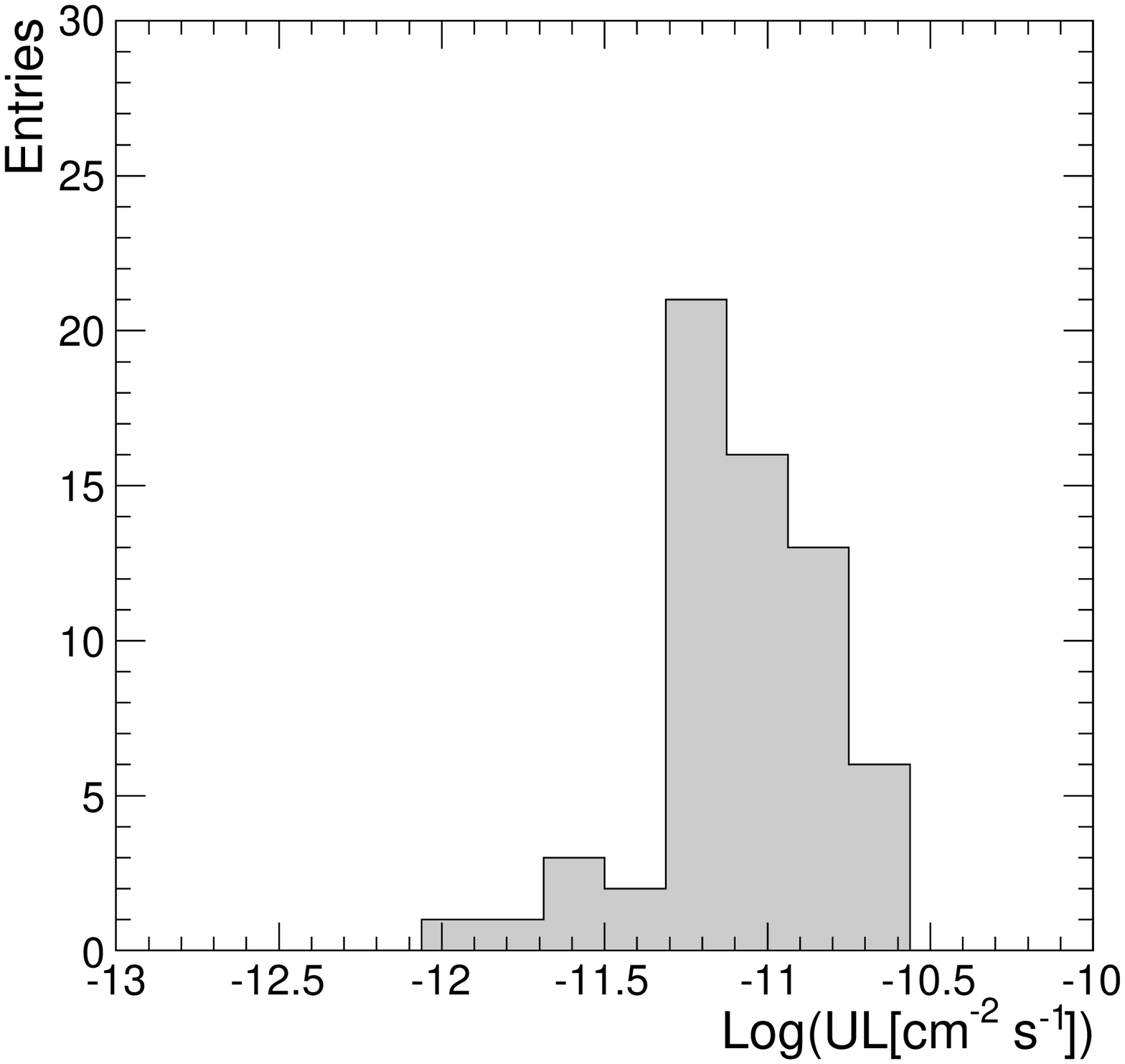}
\caption{\textit{Top}: significance distribution of the sources included in our sample, classified according to their redshift. Sources with unknown z are in blue, sources with $z>0.6$ are in red and sources with $z<0.6$ are in grey. The gaussian function represents the expectation from a randomly distributed sample, with mean equal to zero, and variance equal to 1. \textit{Middle}: same as Top, but for sources classified according to the AGN type. Sy1 galaxies are in green, unidentified sources are in blue, LSP/ISP in red and HSP in grey. \textit{Bottom}: distribution of the values of the integral upper limits, estimated above the correspondent energy thresholds. \textit{Left}: results from the \textit{medium}-cut analysis. \textit{Right}: results from the \textit{soft}-cut analysis.} \label{figone}
\end{figure*} 

\section{\veritas\ observations}
\veritas\ is an array of four 12-m diameter atmospheric Cherenkov telescopes, located at the Fred Lawrence Whipple Observatory (FLWO) in southern Arizona (31 40N, 110 57W,  1.3km a.s.l.). It is sensitive to $\gamma$-rays with an energy above 85 GeV up to $30$ TeV, with an energy resolution of $15$-$20\%$ and an angular resolution $R_{68\%} < 0.1^\circ$. The field of view of \veritas\ is roughly $3.5^\circ$. For more details see \cite{Holder11}.\\

The interaction of the $\gamma$-ray with the Earth's atmosphere triggers a particle shower, which emits Cherenkov photons. The images of the Cherenkov flashes are used to determine the direction and energy of the incoming photon. In the data analysis, different cuts are applied to the parameters of the Cherenkov images to differentiate between $\gamma$-ray-induced showers, and cosmic-ray-induced showers, which represent the main background. The cuts are optimized for sources with different spectral indexes. For our analysis we have used \textit{medium} and \textit{soft} cuts, optimized for sources with an index $\Gamma= -2.4$ and $-4.0$, respectively (see e.g. \cite{Acciari08}). \\

The observations have been performed using the standard \textit{wobble} observation configuration, where the telescopes are pointed $0.5^\circ$ away from the source to allow a simultaneous estimation of the background.\\

The results presented in the next Section have been cross-checked using an independent analysis.

\section{Results}

For each observed source we computed the significance of the detection for both \textit{medium} and \textit{soft} cuts. Given that none of the sources is detected ($\sigma<5$, and indeed all the sources have $\sigma<4$), for each source we computed  the correspondent upper limit on the VHE flux as well. Upper limits are computed at the $99\%$ confidence level, following \cite{Rolke}.\\ 

\subsection{Significance distributions}

In Figure \ref{figone}, we show the significance distributions for all our sources and for both cuts. Above the histograms we plot the gaussian distribution expected from a randomly distributed sample, with mean equal to zero, and variance equal to one. Both distributions deviate from this expectation: the best-fit gaussian functions have a mean of $0.34$ and $0.46$, with variance equal to $1.08$ and $1.26$, for \textit{medium} and \textit{soft} cuts, respectively.\\

We studied two different properties of the sample: the distribution of significances as a function of the distance (top plots of Figure \ref{figone}), and of  the blazar class (medium plots). The major contribution to the positive values of the significance distributions is provided by sources with $z<0.6$, and by HSP blazars, respectively. The first result agrees with the expected absorption of TeV photons by the EBL, which strongly reduces the flux from distant (z$>$0.6) blazars; the latter is consistent with the current population of AGN detected by IACTs, which is dominated by HSP blazars.\\

In the bottom plots of Fig. \ref{figone} we report the distribution of the integral upper limits, estimated above the corresponding energy threshold of each source. For the computation of the upper limits, we assumed an index of $\Gamma=-2.5$ for \textit{medium} cuts, and $\Gamma=-3.5$ for \textit{soft} cuts.\\
The upper limits obtained are, for most of the sources, the first upper limits ever produced at VHE.\\

\subsection{Stacked analysis}

To estimate the significance of the excess that can be seen in Figure \ref{figone}, we performed a stacked analysis of our results. We summed all the excesses observed, and their correspondent uncertainties, computing the significance of the stacked excess. We made this evaluation for the entire data set and for dedicated subsets.\\

The fact that the significance distributions are skewed towards positive values is confirmed by the significance of the stacked excess: for the entire data set, the stacked significance is $\sigma=3.5$, for \textit{medium} cuts, and $\sigma=3.9$, for \textit{soft} cuts, corresponding to an excess of $368$ and $1411$ $\gamma$-rays, respectively.\\
The significance is larger when considering the most favorable of the subsets: HSPs located at $z<0.6$. The stacked analysis on this subset yields $\sigma=3.9$ and $\sigma=4.1$ for \textit{medium} and \textit{soft} cuts, respectively.
On the other hand, if we consider another subset, composed by all sources apart from the HSPs, at a redshift higher than $0.6$, we do not see any excess: in this case we have $\sigma=0.7$ and $\sigma=0.4$ for the two cuts.

\section{Conclusions}

We have presented here the preliminary results from the analysis of 65 extra-galactic sources, observed by \veritas\  from 2007 to 2012. The sample includes mainly blazars, plus several unidentified \fermi\ sources, plus two AGN classified as Seyfert 1 galaxies. None of these sources are detected, all having a significance lower than 4.\\
We have shown the significance distributions of the sample, divided by redshift and by AGN class, as well as the distributions of the integral upper limit values.\\

The significance distributions appear skewed towards positive values when compared to the expectation from a randomly distributed sample. The same trend is shown by a stacked analysis, that has been done for the entire data set, and smaller sub-sets. The entire data set has a stacked significance of $3.5$ and $3.9 \sigma$ for the two cuts used in the analysis. The excess is clearly associated with nearby HSPs, as confirmed by the stacked analysis of the relative data.\\

Non-detection of VHE emission from a candidate source, and the estimation of the correspondent upper limit on its flux, is important for several reasons. Firstly, upper limits can constrain the modeling of the $\gamma$-ray emission, especially for sources detected by \fermi\ at lower energies, but not seen by Cherenkov telescopes.  Secondly, given the extreme variability observed in blazars, if one of these sources will be detected in the future during a flaring state, previous upper limits are fundamental to constrain the variability properties of the object. Finally, all these estimations will be useful in the perspective of observations with the next-generation VHE observatory, CTA.\\

\vspace*{0.5cm}
\footnotesize{{\bf Acknowledgment:}{
This research is supported by grants from the U.S. Department of Energy Office of Science, 
the U.S. National Science Foundation and the Smithsonian Institution, by NSERC in Canada, by 
Science Foundation Ireland (SFI 10/RFP/AST2748) and by STFC in the U.K. We acknowledge the excellent
work of the technical support staff at the Fred Lawrence Whipple Observatory and at the collaborating 
institutions in the construction and operation of the instrument. }
}

\end{document}